\documentclass[5p,superscriptaddress]{elsarticle}

\usepackage{lineno,hyperref}
\usepackage{graphicx} % Grafiken
\usepackage{textcomp}
\modulolinenumbers[5]

\biboptions{square,comma,numbers,sort&compress}
\begin{document}
	
	\begin{frontmatter}
		
		\title{The role of minor alloying in the plasticity of bulk metallic glasses}
		
		
		\author[IMP]{Sven Hilke}
		\author[IMP]{Harald R\"osner\corref{correspondingauthor}}
		\ead{rosner@uni-muenster.de}
		\author[IMP]{Gerhard Wilde}
		
		\cortext[correspondingauthor]{Corresponding author}
		
		\address[IMP]{Institut f\"ur Materialphysik, Westf\"alische Wilhelms-Universit\"at M\"unster, Wilhelm-Klemm-Str. 10, 48149 M\"unster, Germany}

		\begin{abstract}
			Micro- or minor alloying of metallic glasses is of technological interest. An originally ductile Pd-based monolithic bulk metallic glass (Pd$_{40}$Ni$_{40}$P$_{20}$) was selectively manipulated by additions of Fe or Co. The alloying effects were extreme, showing either exceptional ductility upon Co addition or immediate catastrophic failure upon Fe addition when tested under uniaxial compression or 3-point bending. The amorphous structure was characterized prior to deformation with respect to its medium-range order (MRO) using variable resolution fluctuation electron microscopy (VR-FEM). We observe striking differences in the MRO between the ductile and brittle metallic glasses, with the ductile glasses exhibiting a rich structural diversity and MRO correlation lengths up to 6\,nm. The MRO heterogeneity seems to enable easier shear banding and hence enhance the deformability.
		\end{abstract}
		
		
	\end{frontmatter}
	
	%\linenumbers
	
	%\section{Introduction}
	Bulk metallic glasses (BMGs) are promising candidates for applications \cite{schroers2010processing,schroers2011thermoplastic,schroers2013bulk,gibson20183d}. However, most BMGs lack ductility. This is particularly so under tension where, at the end of the elastic regime, immediate catastrophic failure predominates \cite{ashby2006metallic}. Progress has been made in recent years in developing monolithic BMGs exhibiting respectable ductility during cold rolling, bending or compression tests \cite{schuh2007mechanical,eckert2007mechanical,scudino2011ductile,okulov2015flash,nollmann2016impact,scudino2018ductile,bian2019controlling,peng2019deformation,kosiba2019modulating}. The key parameter for the plasticity seems to be the structural heterogeneity of the amorphous materials, which can be manipulated for example by micro- or minor alloying \cite{eckert2007mechanical,wang2011plastic,guo2015structural,nollmann2016impact,wu2016designing,hubek2018impact,im2020structural}. However, the structural heterogeneity is still an ill-defined entity although it is frequently referred to in connection with the amorphous structure of monolithic metallic glasses. Thus, understanding the amorphous structure is of importance. For this purpose, the concept of medium-range order (MRO) is used to describe structural correlations in the amorphous state having length scales beyond that of atomic bonding. Different MRO models have been proposed which suggest MRO correlation lengths or sizes in the range of about 1\,nm \cite{Dash2003}, $1-2$\,nm \cite{Gibson2010,zhan2017effect}, $1-3$\,nm \cite{Bogle2007,borisenko2012medium,Cowley2002}, $1-4$\,nm \cite{Bogle2010,lee2009observation,voyles2002fluctuation} or generally larger than $> 0.5$\,nm (short-range order) \cite{biswas2007real,voyles2000fluctuation,voyles2002fluctuatione,yi2010flexible,zhang2016medium,zhang2018vitrification}.
	
	\noindent In this study we address the structure-property relation using a ternary Pd-based BMG (Pd$_{40}$Ni$_{40}$P$_{20}$), which was selectively manipulated by minor additions of Fe or Co. The glassy materials were characterized using variable resolution fluctuation electron microscopy (VR-FEM). FEM is a microscopic technique based on a statistical analysis of the variance $V(|\vec{k}|,R)$ determined from diffracted intensities of nanometer-sized volumes obtained by scanning transmission electron microscopy (STEM) microdiffraction \cite{iwai1999method,voyles1999experimental,voyles2000fluctuation,voyles2002fluctuation,yi2011effect}. The advantage of FEM is that it contains information on the pair-pair correlations from higher order correlation functions and hence on the MRO \cite{treacy1996variable,treacy2005fluctuation,voyles2002fluctuation,voyles2002fluctuatione,hwang2011variable}.
	The normalized variance $V(|\vec{k}|,R)$ of the spatially resolved diffracted intensity $I$ of a nanobeam diffraction pattern (NBDP) is therefore a function of the scattering vector $\vec{k}$ and the coherent spatial resolution $R$:
	\begin{equation}
	V(|\vec{k}|,R) = \frac{\left\langle I^2 (\vec{k},R,\vec{r}) \right\rangle}{\left\langle I (\vec{k},R,\vec{r}) \right\rangle^2}-1
	\label{EQ:NVAR}
	\end{equation}
	where $\left\langle\,\,\right\rangle$ indicates the averaging over different sample positions $\vec{r}$ or volumes, and $R$ denotes the FWHM of the electron probe \cite{yi2010flexible}.
	 Sampling with different parallel coherent probe sizes, $R$, is called VR-FEM \cite{treacy2007probing,treacy2007structural}. It gives insight into the structural ordering length scale and, using either peak height or peak integral, provides a semi-quantitative measure of the MRO volume fraction. Differences in the MRO may then be compared with differences in the properties of the BMGs \cite{treacy1996variable,hwang2011variable,voyles2002fluctuatione,Davani2019}. 
		
	\begin{figure*}[h]
		\includegraphics[width=\textwidth]{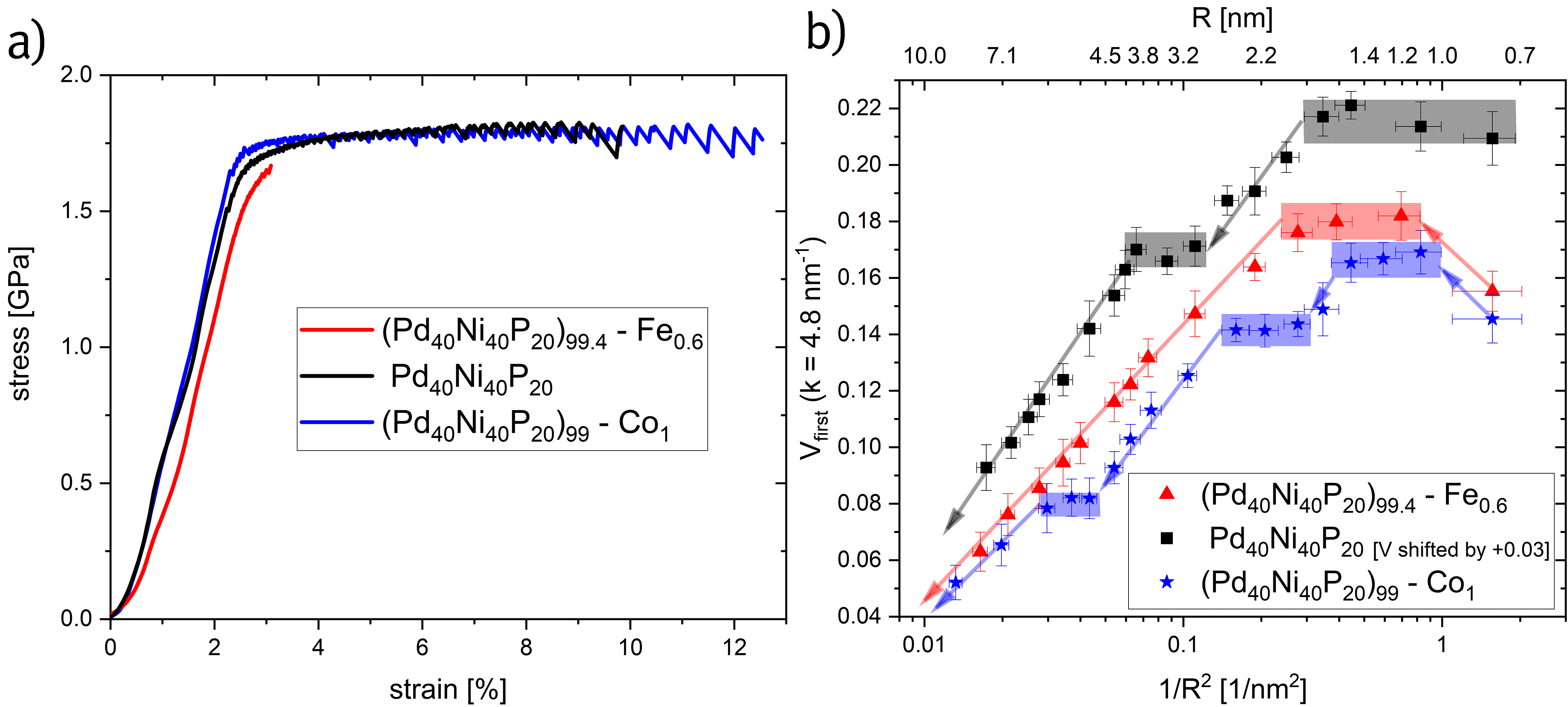}%
		\centering
		\caption{(a) Uniaxial compression tests of Pd$_{40}$Ni$_{40}$P$_{20}$ with additions of Fe and Co. (b) FEM analysis of the corresponding amorphous structure showing V$_{\mathrm{first}}$ at $k = 4.8$\,nm$^{-1}$ plotted against $1/R^2$. For better visibility the curve of the Pd$_{40}$Ni$_{40}$P$_{20}$ master alloy has been shifted upwards ($+0.03$).}
		\label{fig:FIG1}
	\end{figure*}
	
	%\section{Materials and Methods}
	Pd$_{40}$Ni$_{40}$P$_{20}$ samples were produced by melting pure ingots of Pd (99.5\%) and Ni$_2$P (99.9\%) in a melt spinner under argon atmosphere. This master alloy was subsequently manipulated by additions of 1 at.\% Co or 0.6 at.\% Fe. Prior to casting, the ingots were cycled with boron oxide (B$_2$O$_3$) to purify the samples. For the deformation tests samples were cut out of the ingots using a diamond wire saw to give dimensions of 4\,mm (length) x 3\,mm (diameter) providing a 4:3 aspect ratio \cite{conner2003shear,wu2008strength}. Uniaxial compression tests were performed in an Instron (Instron, model 1195) using a strain rate of $2.5 \times 10^{-5}$\,s$^{-1}$. The test device was equipped with home-made anvils of extra hardened B\"ohler S290 microclean steel. Since it has been demonstrated that compression tests to examine plasticity are sensitive to both alignment and shape geometry \cite{wu2008strength}, 3-point bending tests were also performed. For brevity, these results are not included here but they confirmed the results of the compression tests \cite{nollmann2016impact,Davani2019,davani2020shear}. More comprehensive details of the sample processing and deformation are given in reference \cite{horbach2020shear,nollmann2016impact}. Electron-transparent samples were prepared by electropolishing with a BK-2 electrolyte \cite{kestel1986non} at 16.5\,V / $- 20\,^\circ$C using a Tenupol 5 electropolishing device (Struers A/S, Denmark).
	
	\noindent VR-FEM was performed at 300\,kV in a Thermo Fisher Scientific Themis 300 G3 transmission electron microscope (TEM). NBDPs were acquired with parallel coherent probe sizes between 0.8 and 8.5\,nm at FWHM using a 10\,\textmu m C2 aperture. The probe current was set to 15\,pA. The relative foil thickness (t/$\lambda$) of the TEM samples was determined by the log-ratio method using the low-loss part of electron energy loss (EEL) spectra \cite{malis1988eels}. All FEM data presented here were recorded under similar experimental conditions of t/$\lambda$, beam current and acquisition time in order to be sensitive to changes in the MRO measured by the normalized variance signal. The dwell-time was 4\,s for the individual NBDPs acquired with a CCD camera (US 2000) at binning 4 (512x512 pixels). The camera length used was 77\,mm. All probe sizes were measured prior to the FEM experiments directly on the Ceta camera using Digital Micrograph plugins by D. Mitchell \cite{mitchell2005scripting}. FEM analyses were performed on the diffracted intensities $I (\vec{k},R,\vec{r})$ of sets of 100 individual NBDPs taken from the same scanned areas. The normalized variance profiles were calculated using a pixel by pixel analysis according to the annular mean of variance image ($\Omega_{VImage}$(k)) \cite{daulton2010nanobeam,schmidt2015quantitative,gammer2010quantitative}. These are shown in the Supplementary Material (see Figs. S1-S3). MRO volume fractions were estimated by averaging the peak heights of the first variance peak (V$_{\mathrm{first}}$) for each material and subsequently normalizing to that of the Pd$_{40}$Ni$_{40}$P$_{20}$ master alloy (see Tab.~\ref{tab:TAB1}).

	%\section{Results}
	
	\begin{table*}[h]
		\caption{Tabulated data showing the discrete MRO correlation lengths, the relative volume fractions normalized to Pd$_{40}$Ni$_{40}$P$_{20}$ and strain to failure values (averaged) from individual uniaxial compression tests \cite{Nollmann2018}.}
		\centering
		\vspace{5pt}
		\begin{tabular}{c|c|c|c}
			Sample & (Pd$_{40}$Ni$_{40}$P$_{20}$)$_{99.4}$-Fe$_{0.6}$ & Pd$_{40}$Ni$_{40}$P$_{20}$ & (Pd$_{40}$Ni$_{40}$P$_{20}$)$_{99}$-Co$_1$ \\
			\hline \hline
			& & $(1.3\pm0.6)$ & $(1.3\pm0.3)$ \\
			MRO correlation length(s) [nm] & $(1.6\pm0.5)$ & $(3.5\pm0.6)$ & $(2.2\pm0.5)$ \\ 
			& & & $(5.3\pm0.7)$ \\
			\hline
			%MRO volume fraction [integral] & $\Delta = -(1.6\pm0.)\,\%$ & 1 & $\Delta = -(11.7\pm0.)\,\%$ \\ 
			%normalized to Pd$_{40}$Ni$_{40}$P$_{20}$ & & \\
			%\hline 
			MRO volume fraction (V$_{\mathrm{first}}$) & $\Delta = -(4.4\pm0.2)$ & 1 & $\Delta = -(11.6\pm0.7)$ \\
			normalized to Pd$_{40}$Ni$_{40}$P$_{20}$ [\%] & & \\
			\hline
			Maximum plastic strain & $(1.3\pm0.4)$ & $(5.4\pm1.6)$ & $(7.3\pm2.0)$ \\ 
			from compression tests [\%] & averaged over 6 samples  & averaged over 6 samples  & averaged over 5 samples \\
			
		\end{tabular} 
		\label{tab:TAB1}
	\end{table*}
	
	Results from previous deformation tests \cite{nollmann2016impact} are revisited and shown in Fig.~\ref{fig:FIG1}a. The Pd$_{40}$Ni$_{40}$P$_{20}$ master alloy itself shows a remarkable ductility of about 8\,\%  plastic strain. A plastic strain to failure of about 10\,\% was achieved with Co addition (1 at.\,\%), while Fe addition (0.6 at.\,\%) led to failure shortly after reaching the elastic limit. A compilation is given in Tab.~\ref{tab:TAB1}. 
	
	\noindent In the following, our main focus is laid on the glassy structure in order to elucidate factors governing the deformation behavior of metallic glasses. ''Classical diffraction`` analysis (X-ray diffraction or selected area electron diffraction (SAED)) confirmed the amorphous nature of the glasses and revealed no difference between the investigated materials that would explain the large difference in deformability (see Fig. S4 and reference \cite{nollmann2016impact}). The results of the FEM analyses are displayed in Fig.~\ref{fig:FIG1}b in the form of Stratton-Voyles plots \cite{stratton2008phenomenological}, in which the peak intensity of the first normalized variance peak $V(k)$ at $k=4.8\,$nm$^{-1}$ is plotted against $1/R^2$ \cite{stratton2007comparison}. As expected, one sees a general decrease of V(k) as R increases. Superimposed on this are peaks or plateaus occurring when the probe size matches MRO correlation lengths present as a high volume fraction. The less well defined the MRO correlation length, the wider the plateau. The correlation lengths were determined using the maximum and minimum data points in the plateau to calculate an arithmetic mean. The plateaus are indicated by the coloured boxes in Fig.~\ref{fig:FIG1}b. The curve of the (PdNiP)-Fe alloy exhibits a single plateau, yielding an average MRO correlation length of $(1.6\pm0.5)$\,nm. The Stratton-Voyles plot for the very ductile (PdNiP)-Co sample, however, shows three distinct plateaus indicating a much more diverse MRO distribution with the following MRO correlation lengths: $(1.3\pm0.3)$\,nm, $(2.2\pm0.5)$\,nm and $(5.3\pm0.7)$\,nm. The PdNiP ternary master alloy displays two distinct plateaus at $(1.3\pm0.6)$\,nm and $(3.5\pm0.6)$\,nm. These results were confirmed by further data sets taken from different regions in the three samples. Thus, a pronounced heterogeneity in terms of multiple and larger MRO correlation lengths was observed for the ductile materials which was not present in the brittle material. The estimated MRO volume fractions reveal decreases of about 12 $\%$ for Co addition and 4\, $\%$ for Fe addition relative to the Pd$_{40}$Ni$_{40}$P$_{20}$ master alloy. All key values of the FEM analyses are listed in Tab.~\ref{tab:TAB1} and these are critically reviewed below.  
	
	%\section{Discussion}
		
	First of all, it is worth noting that the Poisson$'$s ratio of the three BMGs investigated (here $\nu= 0.4$) remained unaffected by the minor alloying \cite{nollmann2016impact}. Thus, an explanation for good deformability based on the relatively high Poisson$'$s ratio can be excluded here \cite{lewandowski2005intrinsic}.
	
	\noindent Next, we discuss the observed MRO correlation lengths extending up to 6\,nm. Our understanding of MRO is that it describes ordering in terms of a correlation of structural motifs present in the glassy solid at length scales beyond the first neighbour (short-range order) \cite{zhang2018vitrification,ma2009power}. That is, the spatial correlations arise from connected clusters which may serve later as seeds for crystallization. MRO manifests itself in the form of speckles in the diffraction pattern \cite{treacy1996variable,treacy2007structural}. Since the inspection of the individual NBDPs ($\sim$\,6000 in total) revealed only speckle contrast and no diffraction spots (reflections), we exclude the presence of crystalline fractions in the form of distinct nanocrystals for the investigated materials. Thus, the present FEM analyses display only MRO. Experimentally reported MRO correlation lengths in the literature are typically $< 2$\,nm. Thus, our observation of lengths up to about 6\,nm may at first appear surprising. However, firstly it is clear from our data, that not all BMGs contain larger correlation lengths. Secondly, in order to detect larger correlation lengths one must probe with larger probe sizes. Thirdly, different correlation lengths may be present depending on the thermo-mechanical history. Further, as stated in the introduction, some models suggest MRO correlation lengths extending beyond the 2\,nm range. Here we show for the first time MRO correlation lengths extending up to 6\,nm and thus experimentally confirm MRO correlation lengths $> 2$\,nm suggested by various models \cite{Bogle2007,borisenko2012medium,Cowley2002,Bogle2010,lee2009observation,voyles2002fluctuation}. 
	
	\noindent Next we address the question how the plasticity is affected by the MRO volume fraction present in the three materials. A comparison between the relative MRO volume fractions and the mechanical behavior shows no correlation (see Tab.~\ref{tab:TAB1}) since the MRO volume fractions decreased in both modified alloys. Thus, there is no trend apparent to explain the observed mechanical behaviour. Further, the ductility is not explained by the presence of larger MRO correlation lengths in the ductile BMGs, because a recent study \cite{im2020structural} showed that simply increasing MRO size led to embrittlement. Thus, it must be the diversity of the MRO (heterogeneity) which leads to improved ductility.    
		
	\noindent Finally, we address the question how the MRO relates to shear banding in BMGs. Upon inhomogeneous deformation, that is at low temperatures and high stresses, metallic glasses exhibit plasticity in the form of a macroscopic sliding along localized regions called shear bands having thicknesses of about 15 nm or less \cite{donovan1981structure,hieronymus2017shear,hilke2019influence}. The deformed samples which were ductile revealed a high number of finely dispersed shear bands penetrating through the surfaces \cite{hubek2020intrinsic}, while the brittle Fe-doped material developed only a few shear bands of which one of them led to catastrophic failure \cite{hubek2020intrinsic,nollmann2016impact}. Thus, the plasticity of metallic glasses is based upon their ability to form multiple shear bands \cite{schroers2004ductile}. Our observations strongly suggest that the ductility is related to the structural heterogeneity in terms of the diversity of MRO enabling easier shear banding.

	%\section{Conclusions}
	In conclusion: While X-ray diffraction or SAED revealed no difference that explains the difference in deformability, VR-FEM presents clear evidence that minor alloying has a huge impact on the correlation of the glassy structures beyond the typically reported 2\,nm range, which directly affects the ductility of the material. This, in turn, makes minor alloying very promising for tuning the properties of metallic glasses via MRO engineering.

	\section*{Acknowledgements}
	We gratefully acknowledge financial support by the DFG via WI 1899/29-1 (project number 325408982). The DFG is further acknowledged for funding our TEM equipment via the Major Research Instrumentation Program under INST 211/719-1 FUGG. We are also indebted to A. Hassanpour, Drs V. Hieronymus-Schmidt and N. Nollmann for TEM sample preparation and deformation tests, respectively.
	
	%\section*{References}
	
	\bibliography{mybibfile}
	
\end{document}